\documentclass[twocolumn]{aastex7}

\input{Preamble/packages}

\newcommand{\titlename}{A generalized method to measure the Lorentz factor from gamma-ray burst photospheric emission}

\newcommand{\KTH}{KTH Royal Institute of Technology, Department of Physics, SE-10691 Stockholm, Sweden}
\newcommand{\OKC}{The Oskar Klein Centre for Cosmoparticle Physics, AlbaNova University Centre, SE-10691 Stockholm, Sweden}
\newcommand{\IAP}{Sorbonne Universit\'e, CNRS, UMR 7095, Institut d’Astrophysique de Paris (IAP), 98 bis boulevard Arago, 75014 Paris, France}

\newcommand{\cm}[1]{{ \rm{\, cm^{#1}} }} 
\newcommand{\s}[1]{{ \rm{\, s^{#1}} }} 
\newcommand{\keV}[1]{{ \rm{\, keV^{#1}} }} 
\newcommand{\erg}[1]{{ \rm{\, erg^{#1}} }} 
\newcommand{\eflux}{{ \rm{\, erg \, cm^{-2} \, s^{-1}} }} 

\newcommand{\rph}[1]{r_{ \rm{ph}#1 }} 
\newcommand{\upscatter}[1]{\psi_{ \rm{sc}#1 }}
\newcommand{\tcurv}[1]{t_{ \rm{curv}#1 }}
\newcommand{\tvar}[1]{t_{ \rm{var}#1 }} 
\newcommand{\Fob}[1]{F_{ \rm{ob}#1 }} 
\newcommand{\Nob}[1]{N_{ \rm{ob}#1 }} 
\newcommand{\Tob}[1]{T_{ \rm{ob}#1 }} 
\newcommand{\TBB}[1]{T_{ \rm{BB}#1 }} 
\newcommand{\TW}[1]{T_{ \rm{W}#1 }} 
\newcommand{\Eb}[1]{E_{ \rm{b}#1 }} 
\newcommand{\kB}[1]{k_{ \rm{B}#1 }} 
\newcommand{\eff}[1]{\epsilon_{ \gamma#1 }}

\newcommand{\lr}[1]{ \left( #1 \right) } 
\newcommand{\lrb}[1]{ \left[ #1 \right] } 

\begin{document}

\title{\titlename}

\author[0009-0006-9732-052X]{Oscar Wistemar}
\affiliation{\KTH}
\affiliation{\OKC}
\email[show]{wistemar@kth.se}

\author[0000-0002-9769-8016]{Felix Ryde}
\affiliation{\KTH}
\affiliation{\OKC}
\email{fryde@kth.se}

\author[0000-0001-7414-5884]{Filip Alamaa}
\affiliation{\KTH}
\affiliation{\OKC}
\affiliation{\IAP}
\email{filipsam@kth.se}

\shortauthors{Wistemar, Ryde, Alamaa}
\shorttitle{Lorentz factors from photospheric GRB emission}

\correspondingauthor{Oscar Wistemar}

\begin{abstract}
    The properties of gamma-ray bursts (GRBs) that are inferred from observations depend on the value of the bulk Lorentz factor, $\Gamma$. Consequently, accurately estimating it is an important aim. In this work, we present a method of measuring $\Gamma$ based on observed photospheric emission, which can also be used for highly dissipative flows that may lead to non-thermal spectral shapes. For the method to be applicable, two conditions need to be met: the photon number should be conserved in the later stages of the jet, and the original photon temperature must be inferred from the data. The case of dissipation via subphotospheric shocks is discussed in detail, and we show that the method is particularly efficient when a low-energy spectral break is identified. We demonstrate the capabilities of the method by applying it to two different GRB spectra. From one of the spectra, we obtain a value for $\Gamma$ with statistical uncertainties of only $\sim 15$\%, while for the other spectrum we only obtain an upper limit.
\end{abstract}

\section{Introduction}\label{sec:intro}

The spectrum observed from the photosphere in a gamma-ray burst (GRB) comes in all shapes and sizes. Historically, the emission was predicted to have a quasi-thermal profile \citep{Goodman86, Paczynski86}, but even though a thermal spectral component can dominate the overall emission \citep[e.g., ][]{Ghirlanda03, Ryde04, Larsson15, Deng22, Chen24}, as in the exceptional case of GRB 090902B \citep{Abdo09, Ryde10}, these instances are quite rare. However, GRB photospheres can also produce non-thermal spectral shapes, particularly when substantial energy dissipation occurs beneath the photosphere \citep{Rees05, Peer06, Giannios12}. A likely source of such dissipation is radiation-mediated shocks \citep[RMSs;][]{Levinson08, Bromberg11, Levinson12, Beloborodov17, Samuelsson22}, which can significantly broaden an initially narrow thermal photon-distribution. \citet{Samuelsson23} demonstrated that RMS-dominated photospheric spectra are capable of reproducing a wide range of spectral shapes observed in GRBs, suggesting that photospheric radiation may be a plausible explanation for GRB prompt emission.

For quasi-blackbody emission, the relationship between energy flux and temperature of the thermal photon field provides a method to infer physical parameters of the outflow, such as the bulk Lorentz factor and the emission radius \citep{Peer07, Begue14, Vereshchagin20, Zhang21}. Consequently, the identification of thermal features within GRB spectra has enabled constraints on the physical properties of the outflows in many GRBs \citep[e.g., ][]{Ryde09, Larsson15, Wang22, Li23}.

In contrast, for dissipative flows the thermodynamic equilibrium is destroyed and the photon field may not have enough time to re-establish a thermal equilibrium before the photons are released at the photosphere \citep{Beloborodov13}. In such a case, the non-thermal spectrum instead contains signatures of the underlying dissipation processes \citep{Ahlgren15, Samuelsson22}. However, the dissipation may not completely remove all signatures of the original photon field. In an RMS for example, some of the photons are advected through the shock without any significant energy gain \citep{Ito18,Lundman18}. The energy of these photons are only increased due to the shock compression, the total effect of which can be calculated by solving the Rankine-Hugoniot shock jump conditions \citep{Blandford81}. Therefore, the original temperature can be deduced even after the radiation has passed through the shock.

In this paper, we generalize the method that relates the observed spectrum to the physical properties of the outflow \citep{Peer07}, making it applicable to both thermal and non-thermal photospheric spectra. The method can be used provided that the photon number is conserved in the later stages of the jet and that the original photon temperature can be determined from the data, and we specifically show that these conditions are satisfied for RMSs. We apply the method to two GRBs, where we fit one time-resolved spectrum each. We perform the fit with an RMS model \citep[the KRA model;][]{Samuelsson22}, which provides the upstream temperature and from which we can infer the Lorentz factor. Lastly, we show how the observed value of the temperature is related to an additional spectral break below the main peak, that has been identified in several GRBs by using empirical spectral models \citep[e.g.,][see also \citet{Samuelsson23}]{Strohmayer98, Ravasio18}.

The outline of the paper is as follows: in \S \ref{sec:method} we present the equations used to measure the Lorentz factor, in \S \ref{sec:RMS} we explain how they relate to RMSs, in \S \ref{sec:example} we apply the method to two spectra and measure the bulk outflow  Lorentz factors, in \S \ref{sec:break} we discuss how an observed low-energy break can be used, and finally in \S \ref{sec:discussion} we discuss and conclude.

\section{\texorpdfstring{Measurement of $\Gamma$ from photospheric emission}{Measurement of the Lorentz factor from photospheric emission}}\label{sec:method}

In this section, we derive an expression for $\Gamma$ from the properties of the thermal outflow. The method is applicable when the following two conditions are met: (i) the photon number is approximately conserved in the outflow above the Planck radius \citep[the radius where the photons fall out of thermodynamic equilibrium with the plasma,][]{Beloborodov13} and (ii) information about the original blackbody temperature is still present in the observed spectrum.\footnote{Condition i) is likely to apply in GRB jets since Bremsstrahlung and double Compton emission are inefficient above the Planck radius and subphotospheric shocks are expected to be photon rich \citep[][see also \S \ref{sec:RMS}]{Bromberg11, Beloborodov13}.} When these conditions are met, the observed photon flux and temperature can be connected to the photon rate and comoving temperature at some radius $r$, where, due to the properties of a blackbody, the latter two are analytically related.

The central engine frame photon rate, $\mathcal{N}$, at radius $r$ for a relativistic blackbody outflow is given by

\begin{equation}\label{eq:bb_relation}
    \mathcal{N} = \frac{4\pi r^2 \Gamma 4 \sigma {T^{\prime}}^3}{2.7 \kB{}}
\end{equation}

\noindent where $T^\prime$ is the comoving temperature at radius $r$, $\sigma$ is the Stefan-Boltzmann constant, and $\kB{}$ is the Boltzmann constant. The photon rate above is related to the observed photon flux, $\Nob{}$, as

\begin{equation}\label{eq:photon_flux}
    \mathcal{N} = \Nob{} \frac{4\pi d_L^2}{(1 + z)}.
\end{equation}

\noindent Here, $d_L$ is the luminosity distance\footnote{We use values for the cosmology parameters from \citet{Planck20}.}, and $z$ is the redshift. The observed photon flux can be obtained from the observed spectral photon flux, $N_{\rm E}$, as $\Nob{} = \int N_{\rm{E}} dE$.

Using standard assumptions of the comoving density decreasing as $r^{-2}$, the observed temperature, $\Tob{}$, is related to the comoving temperature $T^{\prime}(r)$ as

\begin{equation}\label{eq:T_prime}
    T^\prime(r) = \frac{\lr{1 + z}}{2 \tau^{-2/3} \Gamma} \Tob{}
\end{equation}

\noindent where $\tau$ is the optical depth at radius $r$ (equal to $\tau = \rph{} / r$ with $\rph{}$ being the photospheric radius) and $2\tau^{-2/3}$ accounts for adiabatic cooling and a probabilistic photosphere with emission from various radii and angles \citep{Peer2008, Beloborodov11, Samuelsson23}. By equating Equations~\eqref{eq:bb_relation} and \eqref{eq:photon_flux} and inserting Equation~\eqref{eq:T_prime}, we get $\Gamma$ as a function of the photospheric radius, the observed temperature, and the observed photon flux as

\begin{equation}\label{eq:gamma}
\begin{split}
    \Gamma &= \lr{\frac{\sigma}{2.7 \kB{} 2}}^{1/2} \frac{(1 + z)^2}{d_L} \frac{\rph{} \Tob{}^{3/2}}{\Nob{}^{1/2}} \\
    &= 34 \, (1 +z)^2 \, \frac{T_{\rm{keV}, 1}^{3/2} \, r_{\rm{ph},13}}{d_{L,28} \, \Nob{,2}^{1/2}}
\end{split}
\end{equation}

\noindent where we in the last line used $T_{\rm{keV}, 1} = \kB{} \Tob{} / \lr{10 \keV{}}$, $r_{\rm{ph},13} = \rph{} / \lr{10^{13} \cm{}}$, $d_{L,28} = d_L / \lr{10^{28} \cm{}}$, and $\Nob{,2} = \Nob{} / \lr{100 \s{-1} \cm{-2}}$. We note that Equation~\eqref{eq:gamma} agrees with Equation~(1) in \citet{Peer07}, since the effective size of the emitting region is given by 

\begin{equation}\label{eq:calR}
     \mathscr{R} = \lr{\frac{2.7 \kB{} \Tob{} \Nob{}}{\sigma \Tob{}^4}}^{1/2} = \frac{1}{2^{1/2}} \frac{(1 + z)^2}{d_L} \frac{\rph{}}{\Gamma},
\end{equation}

\noindent where the difference in numerical coefficient depends on our use of the total adiabatic cooling in Equation~\eqref{eq:T_prime}, see also \citet{Peer11}. Below, we give two ways of estimating the photospheric radius, which are used in Equation~\eqref{eq:gamma} to relate observables from the photospheric emission to the Lorentz factor, $\Gamma$.

\subsection{\texorpdfstring{Estimating $\rph{}$ from the observed flux}{Estimating the photospheric radius from the observed flux}}

The Thomson optical depth at radius $r$ is given by $\tau(r) \approx n^\prime \sigma_{\rm T} r / \Gamma$, where $\sigma_{\rm{T}}$ is the Thomson cross section and $n^\prime$ is the comoving lepton number density giving rise to the opacity. In general, $n^\prime$ is given by $n^\prime = n^\prime_e + n^\prime_\pm = n_e^\prime \kappa_\pm$, where $n_e^\prime$ are the electrons associated with the baryons in the outflow, $n_\pm^\prime$ are secondary pairs potentially existing in the flow, and $\kappa_\pm = (n_e^\prime + n_\pm^\prime)/n_e^\prime = (n_e^\prime + n_\pm^\prime)/n_p^\prime$ is the pair multiplicity or, equivalently, the lepton-to-baryon ratio. The comoving lepton density can be found from energy conservation as $n^\prime (r) = L \kappa_\pm / \lr{4\pi r^2 \Gamma^2 m_p c^3}$, where $L$ is the isotropic equivalent luminosity. Therefore, the photospheric radius, where $\tau = 1$, is

\begin{equation}
    \rph{} = \frac{L \sigma_{\rm T} \kappa_\pm}{4 \pi m_p c^3 \Gamma^3}.
\end{equation}

\noindent The luminosity is related to the observed flux as $L = 4\pi d_L^2 \Fob{} / \eff{}$, where $\Fob{}$ is the observed flux. Here, $\eff{}$ is the radiation efficiency, defined as $\eff{} = L_\gamma / L$, where $L_\gamma$ is the isotropic equivalent $\gamma$-ray luminosity. Inserting the photospheric radius into Equation~\eqref{eq:gamma}, we get

\begin{equation}\label{eq:gamma_rph}
    \Gamma = 47 \, (1 +z)^{1/2} \kappa_{\pm}^{1/4} \, \frac{d_{L,28}^{1/4} T_{\rm{keV}, 1}^{3/8} \, \Fob{,-7}^{1/4}}{\eff{,-1}^{1/4} \, \Nob{,2}^{1/8}}
\end{equation}

\noindent with $\Fob{,-7} = \Fob{} / \lr{10^{-7} \eflux{}}$ and $\eff{,-1} = \eff{} / 0.1$. Note that the parameter dependence in Equation~\eqref{eq:gamma_rph} is much weaker compared to Equation~\eqref{eq:gamma}. However, instead of one unknown, $\rph{}$, there are now two unknowns, $\kappa_\pm$ and $\eff{}$. The efficiency might be possible to determine based on afterglow considerations, but the pair multiplicity is typically unknown or assumed to be unity. 


\subsection{\texorpdfstring{Estimating $\rph{}$ from the curvature time}{Estimating the photospheric radius from the curvature time}}\label{sec:t_curv}

Another way to estimate the photospheric radius is to use the observed variability timescale. For a relativistic outflow, the photospheric radius is related to the curvature timescale, $\tcurv{}$, in the observer frame as

\begin{equation}\label{eq:curvature_time}
    \rph{} = \frac{2c\Gamma^2 \tcurv{}}{(1 + z)}.
\end{equation}

\noindent Assuming the observed light curve variability time, $\tvar{}$, corresponds to the curvature timescale, the Lorentz factor can be determined by combining Equations~\eqref{eq:gamma} and \eqref{eq:curvature_time} into

\begin{equation}\label{eq:gamma_tvar}
    \Gamma = 48 \, \frac{1}{1 + z} \, \frac{d_{L,28} \, \Nob{, 2}^{1/2}}{T_{\rm{keV}, 1}^{3/2} \, \tvar{,-1}}
\end{equation}

\noindent where $\tvar{,-1} = \tvar{} / \lr{0.1 \s{}}$. The strength of Equation~\eqref{eq:gamma_tvar} is that the Lorentz factor is determined solely from observable quantities. However, note that it is non-trivial to obtain the curvature time from the light curve (see further discussion on this in \S \ref{sec:discussion}).

\section{Application to radiation-mediated shocks}\label{sec:RMS}

RMSs in GRBs are expected to be photon rich \citep{Bromberg11}. This means that photons produced via Bremsstrahlung and double Compton emission, in the shock and its immediate downstream, are negligible compared to the number of photons advected from the upstream.\footnote{In magnetized RMSs in GRBs, synchrotron emission may alter this picture, see \citet{Lundman19}.} Thus, the photon number remains approximately constant across the shock transition. 

\begin{figure}
    \centering
    \includegraphics[width = \columnwidth]{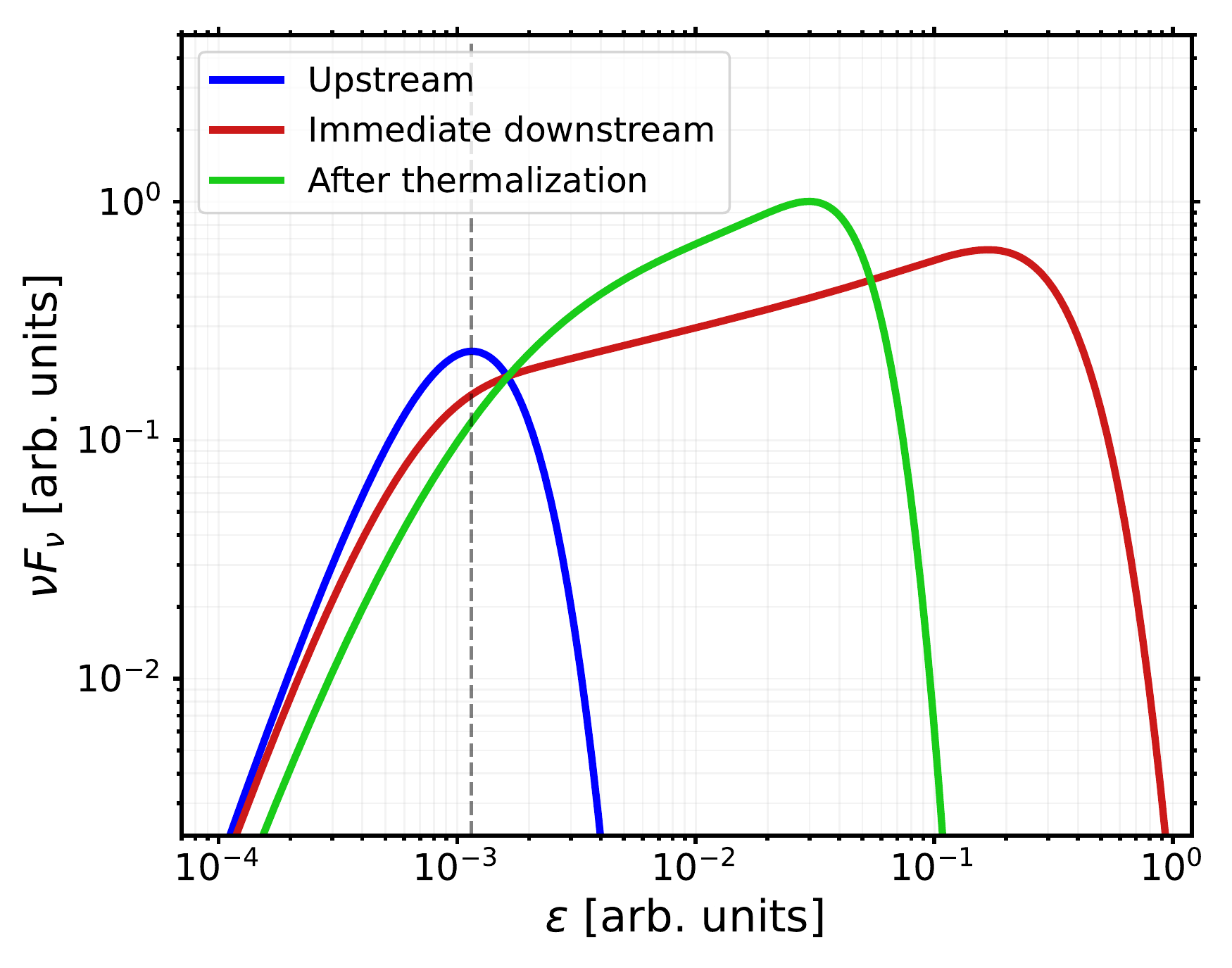}
    \caption{Schematic showing the different stages of the comoving spectrum in the case of dissipation via an RMS. Before the shock, the photon distribution is thermal (blue). In the shock transition region, the photons experience bulk Comptonization leading to a power-law spectrum in the immediate downstream with a low-energy cutoff that is comparable to the the upstream temperature (red). Before the photons are released at the photosphere, they may experience significant thermalization, which shifts the low-energy cutoff to higher energies (green). Adiabatic cooling and shock compression has been neglected for clarity.}
    \label{fig:schematic}
\end{figure}

When photons traverse a photon-rich RMS, they experience bulk Comptonization. Some photons diffuse in the shock for a longer time and gain energy by scattering within the speed gradient of the shock. This generates a power-law for the photon spectral energy distribution in the immediate shock downstream \citep{Ito18, Lundman18}. Since the diffusion process is a random walk, some photons are simply advected through the shock without any significant energy gain from bulk Comptonization. These photons gain energy only due to the adiabatic compression of the plasma \citep{Blandford81}, and they constitute the lowest-energy photons in the immediate downstream from where the power-law extends. In Figure~\ref{fig:schematic}, we show a schematic of how the immediate downstream spectrum relates to the upstream thermal spectrum.

\citet{Samuelsson22, Samuelsson23} showed that photospheric emission that has been energized by RMSs resemble typical prompt GRB spectra. In this picture, the low-energy hardening observed in several GRBs in the X-ray range \citep{Strohmayer98, Ghirlanda07, Guiriec11, Ravasio18} is due to the hard low-energy spectral slope of the adiabatically-compressed, upstream photon-field.

To understand why the method derived in \S \ref{sec:method} is applicable for specifically RMSs, we start by assuming that the upstream plasma is in a thermodynamical equilibrium with temperature $T^{\prime}_u$ (see \S \ref{sec:discussion} for a discussion of this assumption). Accounting for adiabatic compression, the low-energy photons in the downstream can be prescribed a new temperature $T^{\prime}_d = (n_d/n_u)^{1/3} \, T^{\prime}_u$, where $n_d$ ($n_u$) is the downstream (upstream) proper baryon density \citep{Blandford81}. Since the photon number is conserved across the shock transition, the comoving photon density is increased by a factor $(n_d/n_u)$ \citep[e.g.,][]{Ito18}. In Equation~\eqref{eq:gamma}, the Lorentz factor is proportional to $\Tob{}^{3/2}/\Nob{}^{1/2}$, and it is thus independent of the compression.\footnote{Another way to see this is that, since the compression is adiabatic for the low-energy photons, the blackbody relations remain valid.} Therefore, \S \ref{sec:method} is applicable to RMSs as long as the upstream is in a thermodynamical equilibrium.

\section{Measurement of the Lorentz factor in two GRBs}\label{sec:example}

In this section we illustrate how the method can be applied to observed GRB spectra. For this purpose, we have chosen two arbitrary, but typical, GRB spectra. We use data from the Gamma-ray Burst Monitor \citep[GBM;][]{Fermi_GBM} on board the {\it Fermi Gamma-ray Space Telescope}. We choose two spectra that differ in spectral shape and are time-resolved (a small fraction of the total GRB duration) to avoid smearing due to potentially rapid spectral evolution.

In order to do this, we assume that the outflow has been energized by an RMS and use the Kompaneets RMS approximation \citep[KRA;][]{Samuelsson22} to fit the observed spectrum. For data processing we use the Multi-Mission Maximum Likelihood framework \citep[3ML;][]{3ML_Vianello15} and its Bayesian inference setup for analysis. For the Bayesian inference we have used the nested sampler \verb|UltraNest| \citep[]{ultranest}. The observed temperature, energy flux, and photon flux are then given by the posterior.

The two GRBs from which we have selected our two time-resolved spectra have measured redshifts, and we assume $\kappa_\pm = 1$ and $\eff{} = 0.1$. We can then use Equation~\eqref{eq:gamma_rph} to measure the Lorentz factor.

\begin{figure}
    \centering
    \includegraphics[width = \columnwidth]{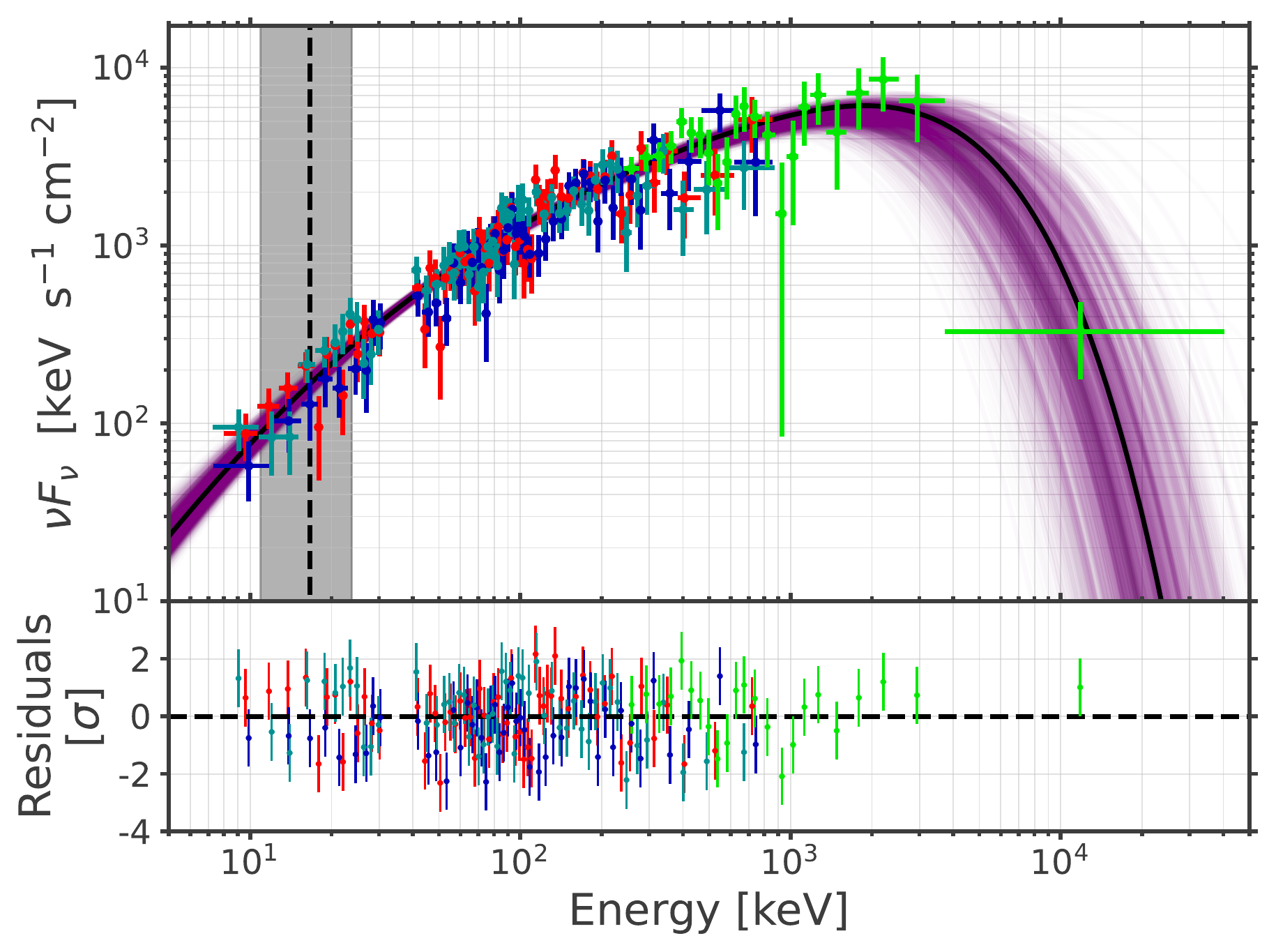}
    \caption{Spectrum used to illustrate the method, here of GRB 130518A at $25.86-26.08 \s{}$ after $T_0$. The purple solid lines show 1500 spectra from the posterior distribution and the black solid line is the best-fit spectrum. The gray shaded region is the $1\sigma$ confidence interval for $\Tob{}$ (multiplied by 3.92, see text for details) and the black dashed line is the best-fit value. The red, blue, teal, and green points are the data points from the NaI3, NaI6, NaI7, and BGO1 detectors, respectively, and represent the best-fit spectrum. Note that data points in a $\nu F_\nu$ representation depend on the model used and their proximity to the model fit should not be over-interpreted and instead evaluated by the residuals.}
    \label{fig:130518}
\end{figure}

GRB 130518A (GBM trigger: 130518580) was detected at 13:54:37.53 UT $(T_0)$ on 18 May 2013 by the GBM \citep{130518_Fermi}. It was also detected by Konus-Wind \citep{130518_KonusWind}. The GBM duration was $T_{90} = 48 \s{} \; (50-300\keV{})$ with a fluence of $9.3 \times 10^{-5} \erg{} / \cm{}^2 \; (10-1000 \keV{})$ and the redshift was measured to be $z = 2.49$ \citep{130518_Fermi, 130518_redshift_osiris, 130518_redshift_gemini}. We use data from four GBM detectors: NaI3, NaI6, NaI7 and BGO1. Figure~\ref{fig:130518} shows a sample of the spectra of the posterior parameter distribution and the best-fit spectrum at $T_0 + 25.86 \s{}$ (time-bin duration $0.22 \s{}$). The bottom panel shows the residuals between the best-fit model and the data. The $68\%$ confidence interval for $\Tob{}$ is also shown.\footnote{$\Tob{}$ has been multiplied by a factor 3.92, since the peak of a Planck spectrum in a $\nu F_\nu$ representation occurs at $3.92 \, \kB{} \Tob{}$.}. The existence of a clear break in the spectrum helps determine the Lorentz factor with a high precision to $\Gamma = 443^{+71}_{-70}$. Note that $\Tob{}$ does not align particularly well with the observed break in the spectrum. Instead, it is slightly shifted towards lower energies compared to the break. A potential reason for this is discussed later in \S \ref{sec:break}.

\begin{figure}
    \centering
    \includegraphics[width = \columnwidth]{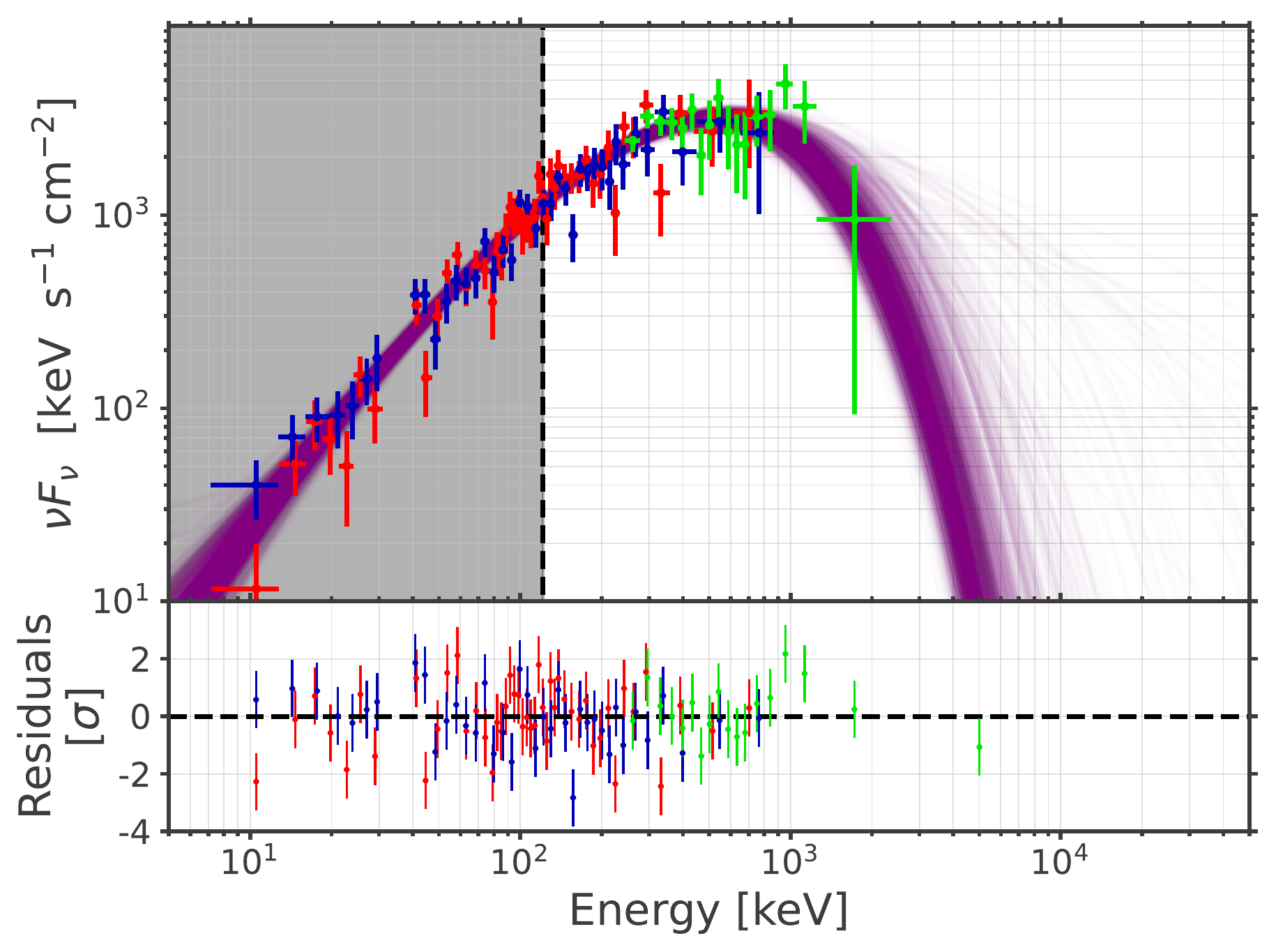}
    \caption{Same as Figure~\ref{fig:130518} but for GRB 210610B at $30.46-30.82 \s{}$ after $T_0$. The gray shaded region shows the possible values for $\Tob{}$ and the black dashed line is the upper limit. The red, blue, and green points are the data points from the NaI9, NaI11, and BGO1 detectors, respectively.}
    \label{fig:210610}
\end{figure}

GRB 210610B (210610827) was detected at 19:51:05.05 UT ($T_0$) on 10 June 2021 by GBM \citep{210610B_Fermi}. It was also detected by multiple other telescopes. The GBM duration was $T_{90} = 55 \s{}$ with a fluence of $9.3 \times 10^{-5} \erg{} / \cm{}^2$ and with a spectroscopic redshift of $z = 1.13$ \citep{210610B_redshift}. We use data from three GBM detectors: NaI9, NaI11 and BGO1. In this case, it is clear from the posterior distribution of $\Tob{}$ that it is unconstrained towards low values of the temperature and, consequently, we can only determine an upper limit. Figure~\ref{fig:210610} shows a sample of the spectra of the posterior distribution at $T_0 + 30.46 \s{}$ (duration $0.36 \s{}$) and the upper limit for the observed temperature.

In both cases the fits are good, which can be confirmed by the randomness of the residuals. In Table~\ref{tab:examples}, the parameters from the fits of both GRBs are listed along with their redshift. For GRB 130518A, $\Tob{}$ is the best-fit value from the posterior parameter distribution, from which we infer the distributions for the energy flux, photon flux, and Lorentz factor. As mentioned above, for GRB 210610B, only an upper limit on $\Tob{}$ can be determined from the posterior. The upper limit is estimated by fitting an exponential cutoff to the high-temperature region of the posterior distribution. The value and $1\sigma$ errors from this estimation are given in the table. Using Equation~\eqref{eq:gamma_rph} with the upper limit (with the errors) on the temperature, yields the upper limit (with errors) on $\Gamma$. On the other hand, since the energy flux and photon flux are largely independent of the temperature, as they are given by the shape of the spectrum, their values are derived from the full posterior distribution.

\begin{deluxetable*}{ccccccD}
\tablecaption{Both GRBs are listed along with their redshift, best-fit values for the time-resolved spectrum, and the measured Lorentz factor. \label{tab:examples}}
\tablewidth{0pt}
\tablehead{
\colhead{GRB} & \colhead{$z$} & \colhead{Time [s]} & \colhead{$\kB{} \Tob{}$ $\lrb{\keV{}}$} & \colhead{$\Fob{}$ $\lrb{10^{-5} \erg{} \s{-1} \cm{-2}}$} & \colhead{$\Nob{}$ $\lrb{\s{-1} \cm{-2}}$} & \colhead{$\Gamma$}}
\startdata
        130518A & $2.49$ & $25.86-26.08$ & $4.2^{+1.8}_{-1.5}$ & $2.7 \pm 0.3$ & $60.5^{+2.8}_{-2.3}$ & $443^{+71}_{-70}$ \\
        210610B & $1.13$ & $30.46-30.82$ & $<30.9 \pm 6.4$ & $1.2 \pm 0.1$ & $32.8^{+1.0}_{-1.1}$ & $<505^{+35}_{-26}$ \\
\enddata
\tablenotetext{}{\textbf{Note:} The Lorentz factor is calculated using Equation~\eqref{eq:gamma_rph}, with $\kappa_\pm = 1$ and $\eff{} = 0.1$. All errors given are for the $68\%$ confidence interval in the posterior distribution. See text for details regarding the upper limits on $\kB{} \Tob{}$ and $\Gamma$.}
\end{deluxetable*}

\section{Observed low-energy spectral break}\label{sec:break}

The method in \S \ref{sec:method} uses the parameter $\Tob{}$, however, as shown in \S \ref{sec:example}, the value of $\Tob{}$ does not necessarily correspond to the observed spectral break. Here, we show how such a spectral break can anyway be used to estimate the Lorentz factor.

As argued in \S \ref{sec:RMS}, the upstream temperature leads to a low-energy break in the immediate downstream spectrum. If one could have observed the immediate downstream spectrum, the break would correspond to $\Tob{}$. However, the electrons in the downstream are at the downstream Compton temperature \citep[][]{Samuelsson22}, which is higher than the temperature of the low-energy photons. Therefore, as the downstream travels towards the photosphere, the low-energy photons may experience significant thermalization, increasing their energy by an average factor $\upscatter{}$ (see Figure~\ref{fig:schematic} for an example spectrum). How large $\upscatter{}$ is depends on the strength of the shock, which affects the value of the Compton temperature, and the optical depth where the dissipation occurs, which affects the total number of scatter events. This thermalization smooths out the break into a curvature, where the break can be more or less defined, and typically increases the break by a factor $\upscatter{}$ (this can be seen in Figure~\ref{fig:130518} where the break is quite smooth and shows a mismatch compared to $\Tob{}$). Therefore, the observed low-energy break, $\Eb{}$, is given by

\begin{equation}\label{eq:E_break}
    \Eb{} = \upscatter{} P \kB{} \Tob{}.
\end{equation}

\noindent where $P$ is introduced to relate the spectral peak of a thermal distribution to $\Tob{}$. The value for $P$ is different depending on if $\Eb{}$ is measured in a photon, energy, or $\nu F_\nu$ spectrum.

Phenomenological models that include a low-energy break in the spectral shape can use their fitted value for the break to estimate $\Tob{}$ using Equation~\eqref{eq:E_break}. One such example is the double smoothly broken power law \citep[DSBPL;][]{Ravasio18}, which has a low-energy break as a free parameter. Another example is a Band function with a subdominant blackbody component \citep{Battelino07, Guiriec11, Siddique22}, in which case the break would correspond to the peak of the blackbody at $\Eb{} = P \, \kB{} \TBB{}$.
Using Equation~\eqref{eq:E_break} together with $P = 3.92$ as appropriate for a blackbody in a $\nu F_\nu$ representation, Equations~\eqref{eq:gamma}, \eqref{eq:gamma_rph}, and \eqref{eq:gamma_tvar} are transformed into
\begin{equation}\label{eq:gamma_ebreak}
    \Gamma = 140 \, \frac{(1 +z)^2}{\upscatter{}^{3/2}} \, \frac{\Eb{,2}^{3/2} \, r_{\rm{ph},13}}{d_{L,28} \, \Nob{,2}^{1/2}},
\end{equation}
\begin{equation}\label{eq:gamma_rph_ebreak}
    \Gamma = 67 \, \frac{(1 +z)^{1/2} \kappa_{\pm}^{1/4}}{\upscatter{}^{3/8}} \, \frac{d_{L,28}^{1/4} \Eb{,2}^{3/8} \, \Fob{,-7}^{1/4} \, }{\eff{-1}^{1/4} \, \Nob{,2}^{1/8}},
\end{equation}
\begin{equation}\label{eq:gamma_tvar_ebreak}
    \Gamma = 12 \, \frac{\upscatter{}^{3/2}}{1 + z} \, \frac{d_{L,28} \, \Nob{,2}^{1/2}}{\Eb{,2}^{3/2} \, \tvar{, -1}},
\end{equation}
\noindent where $\Eb{, 2} = \Eb{} / \lr{10^{2} \keV{}}$.

The introduced upscattering term $\upscatter{}$ is generally unknown. However, one can still use the fact that $\upscatter{} \geq 1$ to obtain an allowed interval for $\Gamma$ using Equations~\eqref{eq:gamma_rph_ebreak} and \eqref{eq:gamma_tvar_ebreak}. As an example, if we assume that the break in Figure~\ref{fig:130518} is at $\Eb{} = 50 \keV{}$, we find $48 \, \tvar{, -1}^{-1} \leq \Gamma \leq 667$ using the values for GRB 130518A given in Table~\ref{tab:examples}. To find the lower limit, a measurement of $\tvar{}$ is required.


\section{Discussion \& Conclusion}\label{sec:discussion}

In this paper, we have outlined a method to measure $\Gamma$ from an observed GRB spectrum. The usefulness of the method depends on how well one can determine the observed temperature, $\Tob{}$. The measurement of $\Tob{}$ is most accurate when a clear low-energy spectral break is present in the data as seen for GRB 130518A in Figure~\ref{fig:130518}. Another example of a burst with a clear low-energy spectral break is GRB 211211A, which has been analyzed in detail using the method presented here (Wistemar, Alamaa, Ryde, 2025, in prep.).

Two situations can make the measurement more difficult. First, as mentioned in \S \ref{sec:break}, thermalization following a dissipation event typically results in a smoothing of the spectrum, potentially hampering the ability to accurately infer the original photon temperature from the data. This in turn can lead to larger uncertainties for the Lorentz factor. Second, when $\Tob{}$ is outside the detector window, the observed low-energy spectral shape may be well described by a single power-law. Thus, only an upper limit on $\Tob{}$ can be set. This leads to an upper or lower limit on the Lorentz factor, depending on which equation is used. In the case of GRB 210610B, we got an upper limit for $\Tob{}$ inside the detector window. Consequently, some of the allowed solutions in the posterior are single power-laws with $\Tob{}$ outside the detector window. 

The discussion above is valid as long as $\Tob{}$ directly corresponds to the original blackbody temperature. However, there exists a region of optical depth where the radiation is still tightly coupled to the plasma but where photon production has become inefficient. If significant dissipation occurs in this zone \citep[called the Wien zone in][]{Beloborodov13}, the radiation forms a Wien distribution at a temperature $\TW{}^{\prime}$, which is higher than the blackbody temperature $\TBB{}^{\prime}$. In this scenario, $\Tob{}$ would most likely correspond to the Wien temperature rather than the blackbody temperature.\footnote{In principle, a Wien and a blackbody temperature could be told apart in the case of exceptional data due to their different spectral shapes. However, in practice this is difficult since the observed emission consists of contributions from various angles and radii, with different comoving temperatures.} One can parameterize this uncertainty by introducing $\psi_{\rm{W}} = \TW{}^{\prime}/\TBB{}^{\prime}$. Then all the equations presented in this paper remain valid as long as $\Tob{}$ is replaced by $\Tob{} / \psi_{\rm{W}}$. We note that when $\Eb{}$ is used, $\psi_{\rm{W}}$ and $\upscatter{}$ have the same dependence and their product forms a combined unknown upscattering factor.

Throughout this paper, we have considered the observed, (non-thermal) prompt emission to be dominated by a single photospheric component. However, in some bursts a clear secondary component is required by the fits during the prompt emission. Prominent examples are given by GRB 090902B \citep{Ryde10} and GRB 1901014C \citep{Ajello2020}. In that case, the observed photon flux, $\Nob{}$, used in \S \ref{sec:method} should correspond to that of the photospheric component only.

In \S \ref{sec:t_curv}, we equated the observed variability time with the curvature time at the photospheric radius. This allowed us to get an expression for $\Gamma$, which was dependent solely on observable quantities. While the curvature timescale is expected to be a characteristic timescale in the observed light curve \citep[e.g., Figure~(5) in][]{Alamaa24}, accurately measuring it can be difficult. Variability time is estimated in several different ways in the literature \citep{MacLachlan13, Scargle13, GolkhouButler14} and the obtained values do not always agree \citep{Yang22, Veres23}. Multiple overlapping pulses emitted from different regions in the jet can further complicate the measurement. Lastly, if the central engine varies on a characteristic timescale, $\delta t_{\rm ce}$, that is longer than $\tcurv{}$, the observed variability time would correspond to $\delta t_{ce}$ instead. Therefore, the determination of $\Gamma$ using Equation~(\ref{eq:gamma_tvar}) must account for systematic uncertainties, particularly those associated with $\tcurv{}$.

In conclusion, we generalized the method to measure the Lorentz factor from \citet{Peer07}, so that it can be used for both thermal and non-thermal photospheric GRB spectra. It is applicable when the photon number is conserved above the Planck radius and when the original photon temperature can be inferred from the data. Then the Lorentz factor is obtained through Equations~\eqref{eq:gamma}, \eqref{eq:gamma_rph}, or \eqref{eq:gamma_tvar}. In \S \ref{sec:RMS}, we argued that these condition apply to RMSs, which produce non-thermal photospheric spectra. We applied the method to two example spectra and used an RMS model \citep[KRA, ][]{Samuelsson22} to make the fits. In the case of GRB 130518A, $\Tob{}$ could be well determined from the data (see Figure~\ref{fig:130518} and Table~\ref{tab:examples}). This leads to an accurate measurement of the Lorentz factor with small statistical uncertainties ($\sim 15$\%). However, for GRB 210610B, the posterior distribution only allowed for an upper limit to be set on $\Tob{}$ (see Figure~\ref{fig:210610}). In such a case, an upper or lower limit can be set on the Lorentz factor, depending on what additional burst information is available. Finally, we noted that identifying a low-energy break in the spectrum allows for the determination of the Lorentz factor using Equations \eqref{eq:gamma_ebreak}, \eqref{eq:gamma_rph_ebreak}, or \eqref{eq:gamma_tvar_ebreak}. Knowledge of the value of the bulk Lorentz factor is of fundamental importance for the understanding of the GRB phenomenon. The method presented in this paper allows for its measurement provided that the emission process is photospheric, even for non-thermal spectra.

\begin{acknowledgments}
    We thank the referee for their insightful comments and suggestions. F.R. acknowledges support from the Swedish National Space Agency (2021-00180 and 2022-00205).  F.A. is supported by the Swedish Research Council (Vetenskapsrådet, 2022-00347). This research has made use of data and/or software provided by the High Energy Astrophysics Science Archive Research Center (HEASARC), which is a service of the Astrophysics Science Division at NASA/GSFC.
\end{acknowledgments}

\bibliography{references}{}

\begin{thebibliography}{}
\makeatletter
\relax
\def\mn@urlcharsother{\let\do\@makeother \do\$\do\&\do\#\do\^\do\_\do\%\do\~}
\def\mn@doi{\begingroup\mn@urlcharsother \@ifnextchar [ {\mn@doi@} {\mn@doi@[]}}
\def\mn@doi@[#1]#2{\def\@tempa{#1}\ifx\@tempa\@empty \href {http://dx.doi.org/#2} {doi:#2}\else \href {http://dx.doi.org/#2} {#1}\fi \endgroup}
\def\mn@eprint#1#2{\mn@eprint@#1:#2::\@nil}
\def\mn@eprint@arXiv#1{\href {http://arxiv.org/abs/#1} {{\tt arXiv:#1}}}
\def\mn@eprint@dblp#1{\href {http://dblp.uni-trier.de/rec/bibtex/#1.xml} {dblp:#1}}
\def\mn@eprint@#1:#2:#3:#4\@nil{\def\@tempa {#1}\def\@tempb {#2}\def\@tempc {#3}\ifx \@tempc \@empty \let \@tempc \@tempb \let \@tempb \@tempa \fi \ifx \@tempb \@empty \def\@tempb {arXiv}\fi \@ifundefined {mn@eprint@\@tempb}{\@tempb:\@tempc}{\expandafter \expandafter \csname mn@eprint@\@tempb\endcsname \expandafter{\@tempc}}}

\bibitem[\protect\citeauthoryear{{Abdo} et~al.,}{{Abdo} et~al.}{2009}]{Abdo09}
{Abdo} A.~A.,  et~al., 2009, \mn@doi [\apjl] {10.1088/0004-637X/706/1/L138}, \href {https://ui.adsabs.harvard.edu/abs/2009ApJ...706L.138A} {706, L138}

\bibitem[\protect\citeauthoryear{{Ahlgren}, {Larsson}, {Nymark}, {Ryde}  \& {Pe'er}}{{Ahlgren} et~al.}{2015}]{Ahlgren15}
{Ahlgren} B.,  {Larsson} J.,  {Nymark} T.,  {Ryde} F.,   {Pe'er} A.,  2015, \mn@doi [\mnras] {10.1093/mnrasl/slv114}, \href {https://ui.adsabs.harvard.edu/abs/2015MNRAS.454L..31A} {454, L31}

\bibitem[\protect\citeauthoryear{{Ajello} et~al.,}{{Ajello} et~al.}{2020}]{Ajello2020}
{Ajello} M.,  et~al., 2020, \mn@doi [\apj] {10.3847/1538-4357/ab5b05}, \href {https://ui.adsabs.harvard.edu/abs/2020ApJ...890....9A} {890, 9}

\bibitem[\protect\citeauthoryear{{Alamaa}}{{Alamaa}}{2024}]{Alamaa24}
{Alamaa} F.,  2024, \mn@doi [\apj] {10.3847/1538-4357/ad5e70}, \href {https://ui.adsabs.harvard.edu/abs/2024ApJ...973...22A} {973, 22}

\bibitem[\protect\citeauthoryear{{Battelino}, {Ryde}, {Omodei}  \& {Band}}{{Battelino} et~al.}{2007}]{Battelino07}
{Battelino} M.,  {Ryde} F.,  {Omodei} N.,   {Band} D.~L.,  2007, in {Ritz} S.,  {Michelson} P.,   {Meegan} C.~A.,  eds,  American Institute of Physics Conference Series Vol. 921, The First GLAST Symposium. AIP, pp 478--479, \mn@doi{10.1063/1.2757410}

\bibitem[\protect\citeauthoryear{{B{\'e}gu{\'e}} \& {Iyyani}}{{B{\'e}gu{\'e}} \& {Iyyani}}{2014}]{Begue14}
{B{\'e}gu{\'e}} D.,  {Iyyani} S.,  2014, \mn@doi [\apj] {10.1088/0004-637X/792/1/42}, \href {https://ui.adsabs.harvard.edu/abs/2014ApJ...792...42B} {792, 42}

\bibitem[\protect\citeauthoryear{{Beloborodov}}{{Beloborodov}}{2011}]{Beloborodov11}
{Beloborodov} A.~M.,  2011, \mn@doi [\apj] {10.1088/0004-637X/737/2/68}, \href {https://ui.adsabs.harvard.edu/abs/2011ApJ...737...68B} {737, 68}

\bibitem[\protect\citeauthoryear{{Beloborodov}}{{Beloborodov}}{2013}]{Beloborodov13}
{Beloborodov} A.~M.,  2013, \mn@doi [\apj] {10.1088/0004-637X/764/2/157}, \href {https://ui.adsabs.harvard.edu/abs/2013ApJ...764..157B} {764, 157}

\bibitem[\protect\citeauthoryear{{Beloborodov}}{{Beloborodov}}{2017}]{Beloborodov17}
{Beloborodov} A.~M.,  2017, \mn@doi [\apj] {10.3847/1538-4357/aa5c8c}, \href {https://ui.adsabs.harvard.edu/abs/2017ApJ...838..125B} {838, 125}

\bibitem[\protect\citeauthoryear{{Blandford} \& {Payne}}{{Blandford} \& {Payne}}{1981}]{Blandford81}
{Blandford} R.~D.,  {Payne} D.~G.,  1981, \mn@doi [\mnras] {10.1093/mnras/194.4.1041}, \href {https://ui.adsabs.harvard.edu/abs/1981MNRAS.194.1041B} {194, 1041}

\bibitem[\protect\citeauthoryear{{Bromberg}, {Mikolitzky}  \& {Levinson}}{{Bromberg} et~al.}{2011}]{Bromberg11}
{Bromberg} O.,  {Mikolitzky} Z.,   {Levinson} A.,  2011, \mn@doi [\apj] {10.1088/0004-637X/733/2/85}, \href {https://ui.adsabs.harvard.edu/abs/2011ApJ...733...85B} {733, 85}

\bibitem[\protect\citeauthoryear{{Buchner}}{{Buchner}}{2021}]{ultranest}
{Buchner} J.,  2021, \mn@doi [The Journal of Open Source Software] {10.21105/joss.03001}, \href {https://ui.adsabs.harvard.edu/abs/2021JOSS....6.3001B} {6, 3001}

\bibitem[\protect\citeauthoryear{{Chen}, {Zhu}, {Peng}  \& {Zhang}}{{Chen} et~al.}{2024}]{Chen24}
{Chen} J.-M.,  {Zhu} K.-R.,  {Peng} Z.-Y.,   {Zhang} L.,  2024, \mn@doi [\apj] {10.3847/1538-4357/ad5f93}, \href {https://ui.adsabs.harvard.edu/abs/2024ApJ...972..132C} {972, 132}

\bibitem[\protect\citeauthoryear{{Cucchiara} \& {Cenko}}{{Cucchiara} \& {Cenko}}{2013}]{130518_redshift_gemini}
{Cucchiara} A.,  {Cenko} S.~B.,  2013, GRB Coordinates Network, \href {https://ui.adsabs.harvard.edu/abs/2013GCN.14687....1C} {14687, 1}

\bibitem[\protect\citeauthoryear{{Deng} et~al.,}{{Deng} et~al.}{2022}]{Deng22}
{Deng} L.-T.,  et~al., 2022, \mn@doi [\apjl] {10.3847/2041-8213/ac8169}, \href {https://ui.adsabs.harvard.edu/abs/2022ApJ...934L..22D} {934, L22}

\bibitem[\protect\citeauthoryear{{Dutta}, {Kumar}, {Sahu}, {Kumar}, {Anupama}, {Bhalerao}, {Barway}  \& {a larger Indian Collaboration}}{{Dutta} et~al.}{2021}]{210610B_redshift}
{Dutta} A.,  {Kumar} H.,  {Sahu} D.~K.,  {Kumar} B.,  {Anupama} G.~C.,  {Bhalerao} V.,  {Barway} S.,   {a larger Indian Collaboration} 2021, GRB Coordinates Network, \href {https://ui.adsabs.harvard.edu/abs/2021GCN.30201....1D} {30201, 1}

\bibitem[\protect\citeauthoryear{{Ghirlanda}, {Celotti}  \& {Ghisellini}}{{Ghirlanda} et~al.}{2003}]{Ghirlanda03}
{Ghirlanda} G.,  {Celotti} A.,   {Ghisellini} G.,  2003, \mn@doi [\aap] {10.1051/0004-6361:20030803}, \href {https://ui.adsabs.harvard.edu/abs/2003A&A...406..879G} {406, 879}

\bibitem[\protect\citeauthoryear{{Ghirlanda}, {Bosnjak}, {Ghisellini}, {Tavecchio}  \& {Firmani}}{{Ghirlanda} et~al.}{2007}]{Ghirlanda07}
{Ghirlanda} G.,  {Bosnjak} Z.,  {Ghisellini} G.,  {Tavecchio} F.,   {Firmani} C.,  2007, \mn@doi [\mnras] {10.1111/j.1365-2966.2007.11890.x}, \href {https://ui.adsabs.harvard.edu/abs/2007MNRAS.379...73G} {379, 73}

\bibitem[\protect\citeauthoryear{{Giannios}}{{Giannios}}{2012}]{Giannios12}
{Giannios} D.,  2012, \mn@doi [\mnras] {10.1111/j.1365-2966.2012.20825.x}, \href {https://ui.adsabs.harvard.edu/abs/2012MNRAS.422.3092G} {422, 3092}

\bibitem[\protect\citeauthoryear{{Golenetskii} et~al.,}{{Golenetskii} et~al.}{2013}]{130518_KonusWind}
{Golenetskii} S.,  et~al., 2013, GRB Coordinates Network, \href {https://ui.adsabs.harvard.edu/abs/2013GCN.14677....1G} {14677, 1}

\bibitem[\protect\citeauthoryear{{Golkhou} \& {Butler}}{{Golkhou} \& {Butler}}{2014}]{GolkhouButler14}
{Golkhou} V.~Z.,  {Butler} N.~R.,  2014, \mn@doi [\apj] {10.1088/0004-637X/787/1/90}, \href {https://ui.adsabs.harvard.edu/abs/2014ApJ...787...90G} {787, 90}

\bibitem[\protect\citeauthoryear{{Goodman}}{{Goodman}}{1986}]{Goodman86}
{Goodman} J.,  1986, \mn@doi [\apjl] {10.1086/184741}, \href {https://ui.adsabs.harvard.edu/abs/1986ApJ...308L..47G} {308, L47}

\bibitem[\protect\citeauthoryear{{Guiriec} et~al.,}{{Guiriec} et~al.}{2011}]{Guiriec11}
{Guiriec} S.,  et~al., 2011, \mn@doi [\apjl] {10.1088/2041-8205/727/2/L33}, \href {https://ui.adsabs.harvard.edu/abs/2011ApJ...727L..33G} {727, L33}

\bibitem[\protect\citeauthoryear{{Ito}, {Levinson}, {Stern}  \& {Nagataki}}{{Ito} et~al.}{2018}]{Ito18}
{Ito} H.,  {Levinson} A.,  {Stern} B.~E.,   {Nagataki} S.,  2018, \mn@doi [\mnras] {10.1093/mnras/stx2722}, \href {https://ui.adsabs.harvard.edu/abs/2018MNRAS.474.2828I} {474, 2828}

\bibitem[\protect\citeauthoryear{{Larsson}, {Racusin}  \& {Burgess}}{{Larsson} et~al.}{2015}]{Larsson15}
{Larsson} J.,  {Racusin} J.~L.,   {Burgess} J.~M.,  2015, \mn@doi [\apjl] {10.1088/2041-8205/800/2/L34}, \href {https://ui.adsabs.harvard.edu/abs/2015ApJ...800L..34L} {800, L34}

\bibitem[\protect\citeauthoryear{{Levinson}}{{Levinson}}{2012}]{Levinson12}
{Levinson} A.,  2012, \mn@doi [\apj] {10.1088/0004-637X/756/2/174}, \href {https://ui.adsabs.harvard.edu/abs/2012ApJ...756..174L} {756, 174}

\bibitem[\protect\citeauthoryear{{Levinson} \& {Bromberg}}{{Levinson} \& {Bromberg}}{2008}]{Levinson08}
{Levinson} A.,  {Bromberg} O.,  2008, \mn@doi [\prl] {10.1103/PhysRevLett.100.131101}, \href {https://ui.adsabs.harvard.edu/abs/2008PhRvL.100m1101L} {100, 131101}

\bibitem[\protect\citeauthoryear{{Li} et~al.,}{{Li} et~al.}{2023}]{Li23}
{Li} L.,  et~al., 2023, \mn@doi [\apjl] {10.3847/2041-8213/acb99d}, \href {https://ui.adsabs.harvard.edu/abs/2023ApJ...944L..57L} {944, L57}

\bibitem[\protect\citeauthoryear{{Lundman} \& {Beloborodov}}{{Lundman} \& {Beloborodov}}{2019}]{Lundman19}
{Lundman} C.,  {Beloborodov} A.~M.,  2019, \mn@doi [\apj] {10.3847/1538-4357/ab229f}, \href {https://ui.adsabs.harvard.edu/abs/2019ApJ...879...83L} {879, 83}

\bibitem[\protect\citeauthoryear{{Lundman}, {Beloborodov}  \& {Vurm}}{{Lundman} et~al.}{2018}]{Lundman18}
{Lundman} C.,  {Beloborodov} A.~M.,   {Vurm} I.,  2018, \mn@doi [\apj] {10.3847/1538-4357/aab9b3}, \href {https://ui.adsabs.harvard.edu/abs/2018ApJ...858....7L} {858, 7}

\bibitem[\protect\citeauthoryear{{MacLachlan} et~al.,}{{MacLachlan} et~al.}{2013}]{MacLachlan13}
{MacLachlan} G.~A.,  et~al., 2013, \mn@doi [\mnras] {10.1093/mnras/stt241}, \href {https://ui.adsabs.harvard.edu/abs/2013MNRAS.432..857M} {432, 857}

\bibitem[\protect\citeauthoryear{{Malacaria}, {Hristov}  \& {Fermi GBM Team}}{{Malacaria} et~al.}{2021}]{210610B_Fermi}
{Malacaria} C.,  {Hristov} B.,   {Fermi GBM Team} 2021, GRB Coordinates Network, \href {https://ui.adsabs.harvard.edu/abs/2021GCN.30199....1M} {30199, 1}

\bibitem[\protect\citeauthoryear{{Meegan} et~al.,}{{Meegan} et~al.}{2009}]{Fermi_GBM}
{Meegan} C.,  et~al., 2009, \mn@doi [\apj] {10.1088/0004-637X/702/1/791}, \href {https://ui.adsabs.harvard.edu/abs/2009ApJ...702..791M} {702, 791}

\bibitem[\protect\citeauthoryear{{Paczynski}}{{Paczynski}}{1986}]{Paczynski86}
{Paczynski} B.,  1986, \mn@doi [\apjl] {10.1086/184740}, \href {https://ui.adsabs.harvard.edu/abs/1986ApJ...308L..43P} {308, L43}

\bibitem[\protect\citeauthoryear{{Pe'er}}{{Pe'er}}{2008}]{Peer2008}
{Pe'er} A.,  2008, \mn@doi [\apj] {10.1086/588136}, \href {https://ui.adsabs.harvard.edu/abs/2008ApJ...682..463P} {682, 463}

\bibitem[\protect\citeauthoryear{{Pe'er} \& {Ryde}}{{Pe'er} \& {Ryde}}{2011}]{Peer11}
{Pe'er} A.,  {Ryde} F.,  2011, \mn@doi [\apj] {10.1088/0004-637X/732/1/49}, \href {https://ui.adsabs.harvard.edu/abs/2011ApJ...732...49P} {732, 49}

\bibitem[\protect\citeauthoryear{{Pe'er}, {M{\'e}sz{\'a}ros}  \& {Rees}}{{Pe'er} et~al.}{2006}]{Peer06}
{Pe'er} A.,  {M{\'e}sz{\'a}ros} P.,   {Rees} M.~J.,  2006, \mn@doi [\apj] {10.1086/501424}, \href {https://ui.adsabs.harvard.edu/abs/2006ApJ...642..995P} {642, 995}

\bibitem[\protect\citeauthoryear{{Pe'er}, {Ryde}, {Wijers}, {M{\'e}sz{\'a}ros}  \& {Rees}}{{Pe'er} et~al.}{2007}]{Peer07}
{Pe'er} A.,  {Ryde} F.,  {Wijers} R. A.~M.~J.,  {M{\'e}sz{\'a}ros} P.,   {Rees} M.~J.,  2007, \mn@doi [\apjl] {10.1086/520534}, \href {https://ui.adsabs.harvard.edu/abs/2007ApJ...664L...1P} {664, L1}

\bibitem[\protect\citeauthoryear{{Planck Collaboration} et~al.,}{{Planck Collaboration} et~al.}{2020}]{Planck20}
{Planck Collaboration} et~al., 2020, \mn@doi [\aap] {10.1051/0004-6361/201833910}, \href {https://ui.adsabs.harvard.edu/abs/2020A&A...641A...6P} {641, A6}

\bibitem[\protect\citeauthoryear{{Ravasio}, {Oganesyan}, {Ghirlanda}, {Nava}, {Ghisellini}, {Pescalli}  \& {Celotti}}{{Ravasio} et~al.}{2018}]{Ravasio18}
{Ravasio} M.~E.,  {Oganesyan} G.,  {Ghirlanda} G.,  {Nava} L.,  {Ghisellini} G.,  {Pescalli} A.,   {Celotti} A.,  2018, \mn@doi [\aap] {10.1051/0004-6361/201732245}, \href {https://ui.adsabs.harvard.edu/abs/2018A&A...613A..16R} {613, A16}

\bibitem[\protect\citeauthoryear{{Rees} \& {M{\'e}sz{\'a}ros}}{{Rees} \& {M{\'e}sz{\'a}ros}}{2005}]{Rees05}
{Rees} M.~J.,  {M{\'e}sz{\'a}ros} P.,  2005, \mn@doi [\apj] {10.1086/430818}, \href {https://ui.adsabs.harvard.edu/abs/2005ApJ...628..847R} {628, 847}

\bibitem[\protect\citeauthoryear{{Ryde}}{{Ryde}}{2004}]{Ryde04}
{Ryde} F.,  2004, \mn@doi [\apj] {10.1086/423782}, \href {https://ui.adsabs.harvard.edu/abs/2004ApJ...614..827R} {614, 827}

\bibitem[\protect\citeauthoryear{{Ryde} \& {Pe'er}}{{Ryde} \& {Pe'er}}{2009}]{Ryde09}
{Ryde} F.,  {Pe'er} A.,  2009, \mn@doi [\apj] {10.1088/0004-637X/702/2/1211}, \href {https://ui.adsabs.harvard.edu/abs/2009ApJ...702.1211R} {702, 1211}

\bibitem[\protect\citeauthoryear{{Ryde} et~al.,}{{Ryde} et~al.}{2010}]{Ryde10}
{Ryde} F.,  et~al., 2010, \mn@doi [\apjl] {10.1088/2041-8205/709/2/L172}, \href {https://ui.adsabs.harvard.edu/abs/2010ApJ...709L.172R} {709, L172}

\bibitem[\protect\citeauthoryear{{Samuelsson} \& {Ryde}}{{Samuelsson} \& {Ryde}}{2023}]{Samuelsson23}
{Samuelsson} F.,  {Ryde} F.,  2023, \mn@doi [\apj] {10.3847/1538-4357/ace441}, \href {https://ui.adsabs.harvard.edu/abs/2023ApJ...956...42S} {956, 42}

\bibitem[\protect\citeauthoryear{{Samuelsson}, {Lundman}  \& {Ryde}}{{Samuelsson} et~al.}{2022}]{Samuelsson22}
{Samuelsson} F.,  {Lundman} C.,   {Ryde} F.,  2022, \mn@doi [\apj] {10.3847/1538-4357/ac332a}, \href {https://ui.adsabs.harvard.edu/abs/2022ApJ...925...65S} {925, 65}

\bibitem[\protect\citeauthoryear{{Sanchez-Ramirez}, {Gorosabel}, {Castro-Tirado}, {Cepa}  \& {Gomez-Velarde}}{{Sanchez-Ramirez} et~al.}{2013}]{130518_redshift_osiris}
{Sanchez-Ramirez} R.,  {Gorosabel} J.,  {Castro-Tirado} A.~J.,  {Cepa} J.,   {Gomez-Velarde} G.,  2013, GRB Coordinates Network, \href {https://ui.adsabs.harvard.edu/abs/2013GCN.14685....1S} {14685, 1}

\bibitem[\protect\citeauthoryear{{Scargle}, {Norris}, {Jackson}  \& {Chiang}}{{Scargle} et~al.}{2013}]{Scargle13}
{Scargle} J.~D.,  {Norris} J.~P.,  {Jackson} B.,   {Chiang} J.,  2013, \mn@doi [\apj] {10.1088/0004-637X/764/2/167}, \href {https://ui.adsabs.harvard.edu/abs/2013ApJ...764..167S} {764, 167}

\bibitem[\protect\citeauthoryear{{Siddique}, {Sajjad}  \& {Motiwala}}{{Siddique} et~al.}{2022}]{Siddique22}
{Siddique} I.,  {Sajjad} S.,   {Motiwala} K.,  2022, \mn@doi [\apj] {10.3847/1538-4357/ac8d05}, \href {https://ui.adsabs.harvard.edu/abs/2022ApJ...938..159S} {938, 159}

\bibitem[\protect\citeauthoryear{{Strohmayer}, {Fenimore}, {Murakami}  \& {Yoshida}}{{Strohmayer} et~al.}{1998}]{Strohmayer98}
{Strohmayer} T.~E.,  {Fenimore} E.~E.,  {Murakami} T.,   {Yoshida} A.,  1998, \mn@doi [\apj] {10.1086/305735}, \href {https://ui.adsabs.harvard.edu/abs/1998ApJ...500..873S} {500, 873}

\bibitem[\protect\citeauthoryear{{Veres} et~al.,}{{Veres} et~al.}{2023}]{Veres23}
{Veres} P.,  et~al., 2023, \mn@doi [\apjl] {10.3847/2041-8213/ace82d}, \href {https://ui.adsabs.harvard.edu/abs/2023ApJ...954L...5V} {954, L5}

\bibitem[\protect\citeauthoryear{{Vereshchagin} \& {Siutsou}}{{Vereshchagin} \& {Siutsou}}{2020}]{Vereshchagin20}
{Vereshchagin} G.~V.,  {Siutsou} I.~A.,  2020, \mn@doi [\mnras] {10.1093/mnras/staa868}, \href {https://ui.adsabs.harvard.edu/abs/2020MNRAS.494.1463V} {494, 1463}

\bibitem[\protect\citeauthoryear{{Vianello} et~al.,}{{Vianello} et~al.}{2015}]{3ML_Vianello15}
{Vianello} G.,  et~al., 2015, \mn@doi [arXiv e-prints] {10.48550/arXiv.1507.08343}, \href {https://ui.adsabs.harvard.edu/abs/2015arXiv150708343V} {p. arXiv:1507.08343}

\bibitem[\protect\citeauthoryear{{Wang}, {Zheng}  \& {Jin}}{{Wang} et~al.}{2022}]{Wang22}
{Wang} Y.,  {Zheng} T.-C.,   {Jin} Z.-P.,  2022, \mn@doi [\apj] {10.3847/1538-4357/aca017}, \href {https://ui.adsabs.harvard.edu/abs/2022ApJ...940..142W} {940, 142}

\bibitem[\protect\citeauthoryear{{Xiong}}{{Xiong}}{2013}]{130518_Fermi}
{Xiong} S.,  2013, GRB Coordinates Network, \href {https://ui.adsabs.harvard.edu/abs/2013GCN.14674....1X} {14674, 1}

\bibitem[\protect\citeauthoryear{{Yang} et~al.,}{{Yang} et~al.}{2022}]{Yang22}
{Yang} J.,  et~al., 2022, \mn@doi [\nat] {10.1038/s41586-022-05403-8}, \href {https://ui.adsabs.harvard.edu/abs/2022Natur.612..232Y} {612, 232}

\bibitem[\protect\citeauthoryear{{Zhang}, {Wang}  \& {Li}}{{Zhang} et~al.}{2021}]{Zhang21}
{Zhang} B.,  {Wang} Y.,   {Li} L.,  2021, \mn@doi [\apjl] {10.3847/2041-8213/abe6ab}, \href {https://ui.adsabs.harvard.edu/abs/2021ApJ...909L...3Z} {909, L3}

\makeatother
\end{thebibliography}
\bibliographystyle{mnras}

\end{document}